# On the mechanism responsible for unconventional thermal behaviour during freezing

Virkeshwar Kumar, G S Abhishek, Atul Srivastava, Shyamprasad Karagadde[*]
Department of Mechanical Engineering, Indian Institute of Technology Bombay, Mumbai 400076, India
[*] Corresponding author: email address - s.karagadde@iitb.ac.in

**Abstract**

In this study, identical experiments of bottom-cooled solidification fluidic mixtures that exhibit faceted and dendritic microstructures were performed. The strength of compositional convection was correlated with the solidifying microstructure morphology, with the help of separate Rayleigh numbers in the mushy and bulk-fluid zones. While the dendritic solidification experienced a monotonic decrease in the bulk fluid temperature, solidification of the faceted case revealed an unconventional, anomalous temperature rise in the bulk liquid, at the initiation of the eutectic phase. Based on the bulk-liquid temperature profile, three distinct regimes of heat transfer were observed in the liquid over the course of solidification, namely - convection-dominated, transition, and conduction-dominated. The observations were analyzed and verified with the help of different initial compositions, as well as other mixtures that form faceted morphology upon freezing. The observed temperature rise was further ascertained by performing an energy balance in an indicative control volume ahead of the solid-liquid interface. The plausible mechanism behind the gain in temperature of the liquid during freezing was further generalized with the help of a simplified one-dimensional numerical model, and was extended to metals which are low Prandtl number mixtures. The study sheds new insights into the role of microstrostructural morphology in governing the transport phenomena in the bulk liquid.

## 1. Introduction

Natural convection in liquid during solidification can considerably affect the composition and temperature distribution, which in turn alters the growth of the solid phase in the mushy zone. The buoyant forces arising due to the rejection of less-dense residual liquid in the mushy zone (e.g., the bottom cooling of the hyper-eutectic water-salt systems) resist the retarding viscous forces [1,2]. The formation of the plumes, which are often termed as solute channels, or chimney [3–5], is known to occur when the strength of the buoyant force exceeds a minimum threshold [2,5].

Likewise, the stabilization of convection in the form of plumes is typically characterized by a critical Rayleigh number in the permeable/mushy zone. Poirier [6] reported the empirical relations of permeability based on the flow directions with respect to the orientation of growing solid. Some popular empirical relations are multilinear regression, Hagen-Poiseuille model, Blake-Kozeny model, slit model, crossflow model. These empirical relations were based on solidification of Pb-Sn, borneol-paraffin systems. Schneider et al. [7] reported the coefficient of Blake-Kozeny equation, where permeability was assumed as isotropic for columnar growth. Felicelli et al. [8] reported the modeling studies for the plume formation and number of empirical relations were introduced for the permeability measurements in the mushy zone with their coefficient. In their work, the permeability relations were categorized on the basis of solid fraction. Worster [9,10] formulated a permeability-based Rayleigh number in the mushy zone using two different types of length scales, i.e., (a) height of the mushy zone and (b) the ratio of thermal diffusivity to the growth rate. Ramirez and Beckerman [11] reported a combined study for the formation of plumes in Pb-Sn and Ni-base superalloys, with the average permeability estimated using both Blake-Kozeny and Poirier's relations and the two prior mentioned length scales. For the Pb-Sn based alloys, the critical permeability based Rayleigh number for plume formation was 38-46, whereas 30-33 for Ni-based superalloys. Yuan and Lee [4] stated that the competition between upward solute transport and dendritic growth determined the survival of the plumes.

The solidifying interface can be classified as either diffuse or faceted, where the formation of the latter is primarily attributed to a higher entropy of phase change at the melting temperature [12]. Dendritic morphology is a diffuse interface which is mostly found in metallic systems and a few transparent materials such as succinonitrile, camphor, $NH_4Cl$, $Na_2ClO_3$, etc [3,13–15]. The faceted growth is generally observed in semiconductors, oxides, carbides, and complex non-metallic compounds such as $KNO_3$, $CuSO_4$, $Na_2SO_4$ [12,14,16–20]. The microstructural morphology together with the solid fraction (equivalent to a packing fraction in a porous medium) considerably affects the transport of rejected components, particularly when it is lighter compared to the bulk fluid density. A number of studies have reported convection and temperature measurements in the liquid during bottom cooled solidification of hyper-eutectic water-$NH_4Cl$ system [21,22], in which the primary convection is compositionally driven, in the presence of a positive thermal gradient in the vertical direction. However, contrasting flow and thermal behaviors have been reported when the primary solid is in the faceted form. Vigorous convective flow was observed during the solidification of water-$Na_2SO_4$, and water-$CuSO_4$ [13,16,20] under side and top cooled conditions. Thompson et al. [23] performed bottom-cooled experiments involving $KNO_3$ as one of the constituents, and reported vigorous and random convection for a certain duration where $KNO_3$ was the primary solidifying component. Further, a gain in the bulk fluid temperature was also observed at the end of convection in [23], which was attributed to the heat gained from the surroundings. This may not be necessarily true, as the distinguishing flow patterns with dendritic and faceted morphologies control the heat and mass transfer within the liquid.

In order to understand the mechanisms of flow under the said two morphological conditions, and to assess the plausible physical reasons that may explain the uncharacteristic temperature rise, we performed real-time observations of bottom-cooled solidification of water-$NH_4Cl$ and water-$KNO_3$ mixtures, with measurement of temperature from several locations along of the height of the solidification cell. The density-based shadowgraph technique and particle image velocimetry (PIV) were employed for performing the qualitative and quantitative flow characterization, respectively. Two different Rayleigh numbers, one defined in the bulk fluid and the other in mushy zone volumes were evaluated to measure the convective strength in the respective zones. The temperature rise during the faceted growth has been quantified, and a plausible hypothesis for the anomalous temperature behaviour is proposed.

## 2. Experimental setup and procedure
### 2.1 Materials

Water-KNO3 and water-$NH_4Cl$ aqueous binary solutions were used to obtain faceted and dendritic morphology of the primary β-solid, respectively. The density of water, $NH_4Cl$, and $KNO_3$ are 1, 1.57, and 2.11g/cm$^3$, respectively. Eutectic temperatures of binary eutectic systems, water-9 wt% $KNO_3$, and water-19.7 wt% $NH_4Cl$ are -2.84 ℃ and -13.9 ℃, respectively [24,25] (representative phase diagrams are shown in Figure 1). At these compositions, the expected solid fraction values at the eutectic temperature are 0.17 and 0.07. In the present study, the hyper-eutectic regime was chosen to perform experiments where a low-density mixture was rejected at the solidifying interface. Water-23 wt% $KNO_3$ and water-24 wt% $NH_4Cl$ were the initial compositions, and their corresponding liquidus temperatures were 18 ℃, and 6 ℃ respectively [24,26]. The details of thermo-physical properties are listed in Table 1. For the verification of flow phenomena in faceted cases, different initial compositions which are near the eutectic composition, such as water-(15 and 20) wt% $KNO_3$ system are discussed in the appendix. Additionally another faceted forming system (water-18 wt% $Na_2SO_4$) is discussed in the appendix.

Table 1: Thermo-physical properties of water-$KNO_3$ and water-$NH_4Cl$ systems [23,24,26–28]

| Parameters (Units) | Symbol | water-$KNO_3$ | water-$NH_4Cl$ |
| --- | --- | --- | --- |
| Initial composition (wt %) | $C_o$ | 23 | 24 |
| Eutectic composition (wt %) | $C_E$ | 9 | 19.7 |
| Eutectic Temperature (°C) | $T_E$ | -2.84 | -13.91 |
| Liquidus Temperature (°C), (C wt% of $KNO_3$ or $NH_4Cl$) | $T_l$ | 1.5C-16.305 | 4.61C-104.6 |
| Thermal diffusivity ($m^2/s$) | $\alpha$ | $1.33 \times 10^{-7}$ | $1.5 \times 10^{-7}$ |
| Solutal diffusivity ($m^2/s$) | $D$ | $2.23 \times 10^{-9}$ | $2.2 \times 10^{-9}$ |
| Thermal conductivity of liquid ($W/m.k$) | $k_l$ | 0.52 | 0.468 |
| Thermal conductivity of solid ($W/m.k$) | $k_s$ | 2.6 | 2.7 |
| Solutal expansion coefficient ($wt\ \%^{-1}$) | $\beta_C$ | $6.04 \times 10^{-3}$ | $2.57 \times 10^{-3}$ |
| Kinematic viscosity ($m^2/s$) | $\nu$ | $1.21 \times 10^{-6}$ | $1.21 \times 10^{-6}$ |
| Density of liquid ($Kg/m^3$) at $T_E$ | $\rho_l$ | 1063 | 1055 |
| Density of solid ($Kg/m^3$) | $\rho_s$ | 1527 | 1527 |
| Latent heat of fusion ($KJ/Kg$) | $H$ | 496 | 313 |

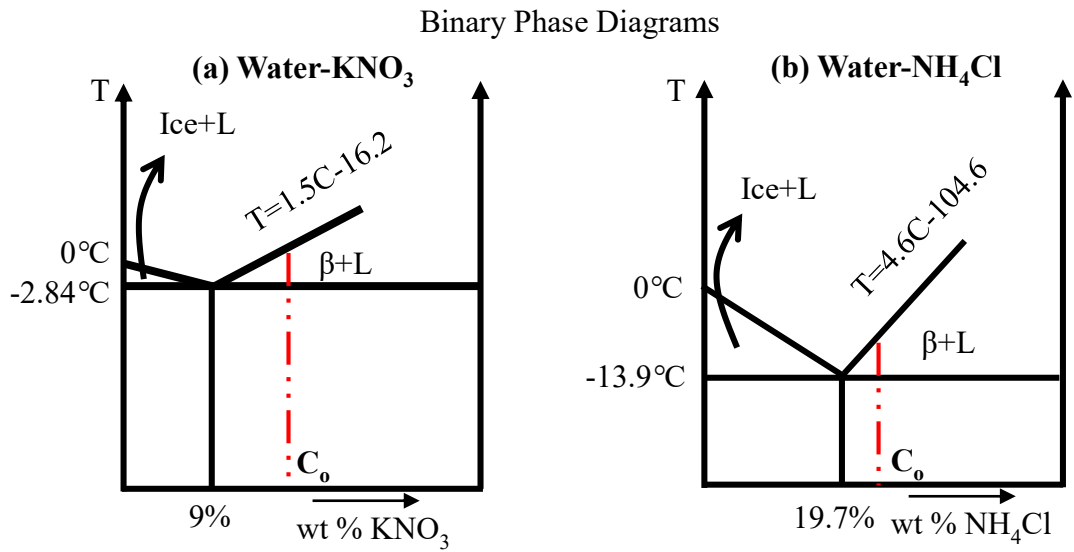

Figure 1: Schematic of the phase diagrams of (a) water-$KNO_3$ (b) water-$NH_4Cl$. $C_o$ (dotted line) is the representative initial composition for experiments.

## 2.2 Experimental setup, instrumentation and measurements

The bottom-cooled solidification cell was made of 8 mm thick perspex glass, as shown in Figure 2. The inner volumetric dimension of the cell was $110 \times 70 \times 50$ mm$^3$. To prevent the condensation problem during *in situ* experiments, a cover cell was placed on the solidification cell. The bottom copper plate of the cell was maintained at -22 °C using an assembly of peltier, TEC controller, DC power supply, heat exchanger, and a julabo circulating bath [22,29]. Nine *K*-type thermocouples (accuracy of ±0.15 °C) were attached to each of the vertical sidewalls to measure the temperature. The temperature data was acquired through data loggers (PICO ® TC-08).

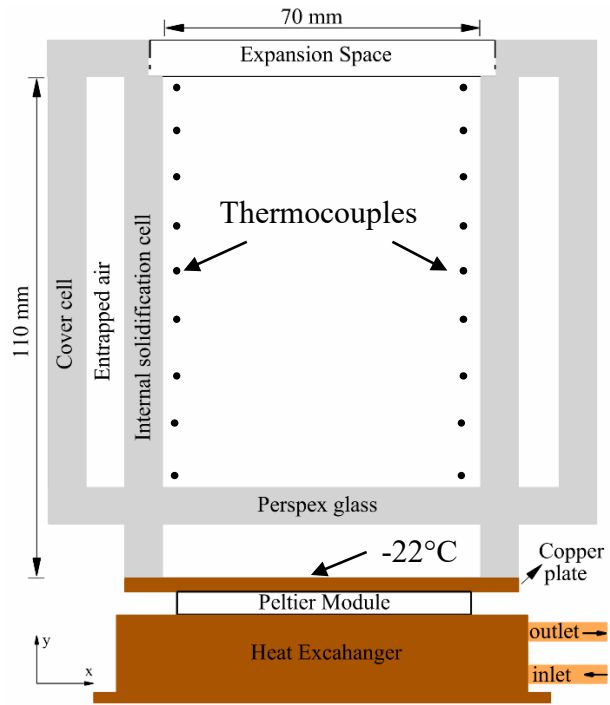

Figure 2: Schematic of the solidification cell.

The density is known to be a function of temperature and composition. For observing the buoyant convection patterns during the experiments, the density gradient based shadowgraph technique was employed. In the shadowgraph, a white light source (250 W) was collimated using a plano-convex lens. The collimated beam was passed through the solidification cell, and density changes in the cell were captured by a CCD camera (Imperx CLB-B2520M). For the velocity measurements of liquid, particle image velocimetry (MicroVec PIV ®) was employed. PIV system consists of a DPSS laser (532 nm), synchronizer, and a CCD camera (Imperx CLB-B2520M, resolution 2456 × 2048). Neutrally buoyant hollow glass particles (average diameter 10 μm) were used as seeding particles. The details of the employed PIV setup can be found in the references [30,31].

The calibration curve of the refractive index with known composition was established using a refractometer (Anton Paar ®). Samples were extracted from the cell during the experiment, and the refractive index was measured. Using the calibration curve and measured refractive index of samples, the composition of $KNO_3$ and $NH_4Cl$ was evaluated.

## 3. Observations

### 3.1 Convective flows in faceted and dendritic growth cases

Figure 3 shows the spatio-temporal evolution of the flow field obtained from the shadowgraphs during bottom-cooled solidification of water-23 wt% $KNO_3$ (a-c), and water-24 wt% $NH_4Cl$ (d-f). In water-23 wt% $KNO_3$, the flow of rejected (low dense) mixture was random (not in specific patterns such as plume, Figure 3(a)), whereas in water-24 wt% $NH_4Cl$, the plume formation and localized solutal convection (near bottom section) were observed (Figure 3(d)). Evolving microstructures were captured using a portable microscope (Dinolite ®, Model AM7515MZT). The faceted and dendritic solid were observed in water-23 wt% $KNO_3$ (inset of Figure 3(b)), water-24 wt% $NH_4Cl$ (inset of Figure 3(e)) respectively.

The random convective flow essentially homogenized the composition and temperature of the liquid in the cell (particularly ahead of the mushy zone) and reduced the bulk composition to near eutectic composition within 300 min. Note that in Figure 3(c), the macroscopic freezing interface is slightly inclined, which could be attributed to slight variations in the compositional distribution induced by the random convective flow patterns. The composition of the bulk fluid and the strength of convection were found to further reduce with time, which was measured to be water-11.8 wt % $KNO_3$ at 350 min. At 600 min, the composition reached the uniform eutectic composition water-9.9±.3 wt% $KNO_3$, and interface became flat (Figure 4(a-c)) from previous time instants (Figure 3(c)).

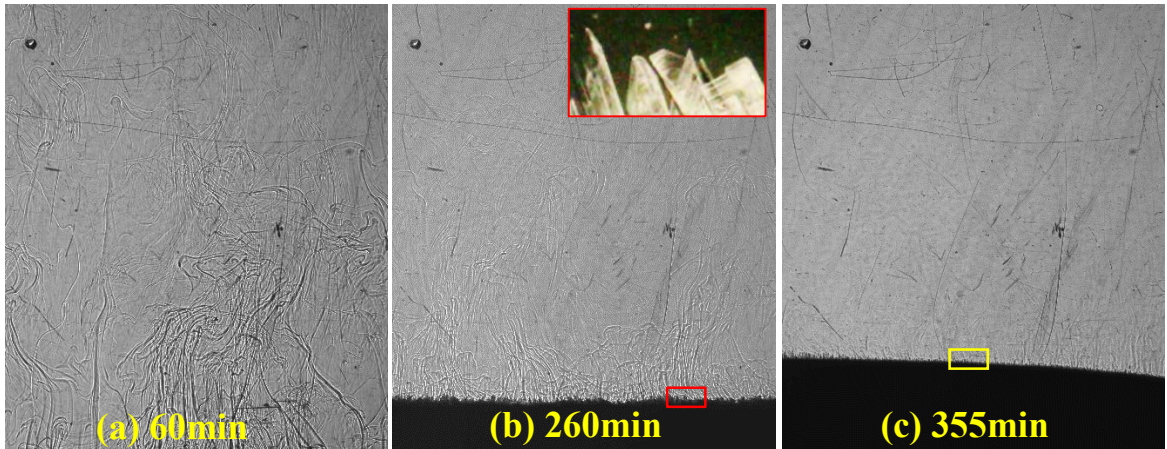
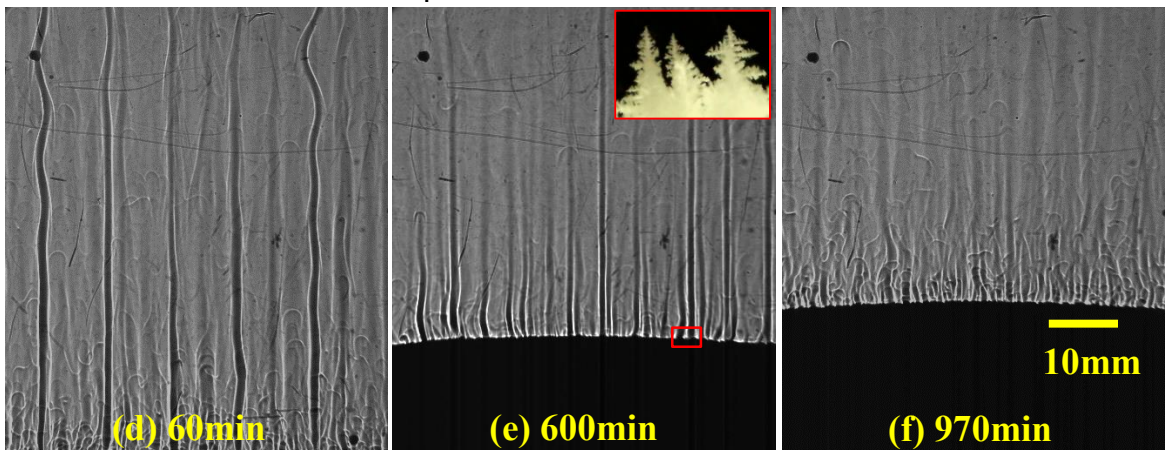

Figure 3: Convective flow observation using shadowgraph (a-c) for water-23 wt% $KNO_3$: (a) random solutal convection at 60 min (b) similar flow at 260 min, and faceted microstructure is shown in the inset of figure, (c) the composition reached near eutectic composition and the sloped interface was observed at 355 min, and (d-f) for water-24 wt% $NH_4Cl$ (d) plume formation at 60 min (e) dendritic morphology captured at solid-liquid interface presented in the inset of figure at 600 min (f) localized solutal convection at 970 min. (Note: The yellow rectangular box on the interface in (c) used for intensity measurements).

In the case of dendritic growth, the solutal convection dominated in the form of plumes, which developed a solutal gradient in the direction of gravity. Hence, it would take a much longer time to reach the eutectic composition. At 600 min, the compositions at the top of the cell and near the solidifying interface were water-21.3 wt% $NH_4Cl$ and water-22.47 wt% $NH_4Cl$ respectively. The growth was comparatively slower between 970 min (Figure 3(f)) to 1410 min (Figure 4(d)) due to continuous change in composition and the liquidus temperature. Composition reached closer to the eutectic composition (water-20.05 wt% $NH_4Cl$) at 1410 min. Furthermore, eutectic solid was observed much later, at 1410 min (Figure 4(e, f)), compared to 600 min in the faceted case.

**Faceted growth (**water-$KNO_3$**)**

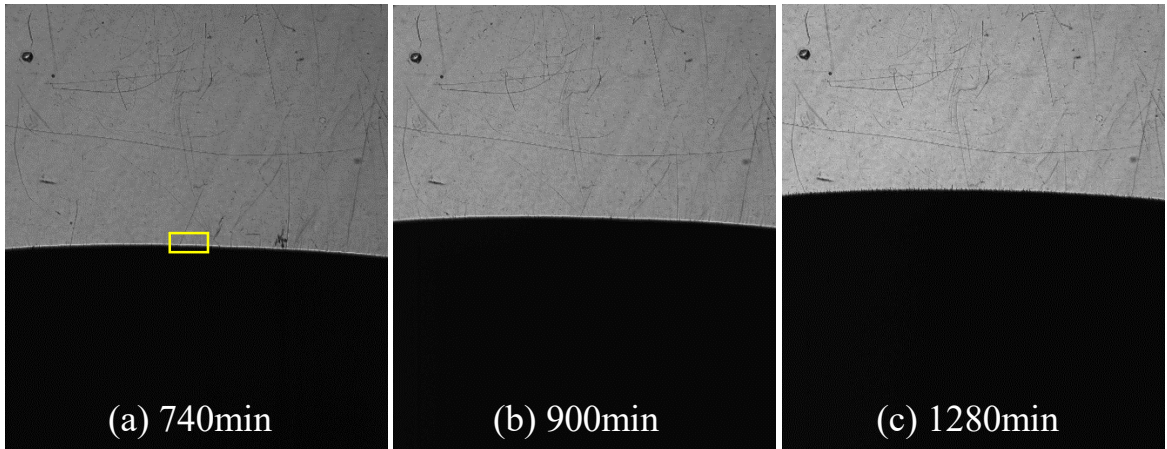

(a) 740min    (b) 900min    (c) 1280min

**Dendritic growth (**water-$NH_4Cl$**)**

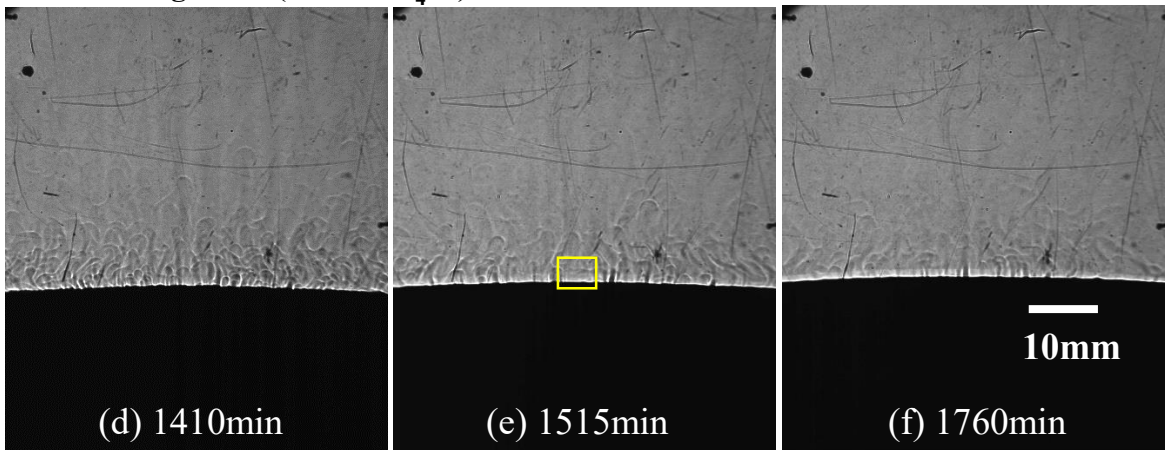

(d) 1410min    (e) 1515min    (f) 1760min

Figure 4: Eutectic growth was observed using shadowgraph (a-c) for water-23wt% $KNO_3$: at (a) 740 min (b) 900 min (c) 1280 min and (d-f) for water-24wt% NH4Cl: (d) localized solutal convection occurs with composition reached near the eutectic composition at 1410 min (e) eutectic growth at 1515 min (f) similar eutectic growth at 1760 min. The yellow rectangular box on the interface in (a, d, and e) used for intensity measurements.

### 3.2 Role of convection on the evolution of solid

In the two hyper-eutectic binary mixtures (water-23 wt% $KNO_3$ and water- 24 wt% $NH_4Cl$), the primary solid (β) phase is primarily the salt system. At temperatures below the eutectic line, the solid exists as a mixture of β-solid and the eutectic (existing in the inter-granular regions). In the present experiments, such a solid phase was observed at the bottom of the cell and is termed as β+eutectic solid (Figure 5(a, b)). On comparing the solidifying structures of the two binary mixtures, the formation of a larger amount of eutectic solid in the faceted case was observed, as the liquid ahead of the freezing mixture was able to reach near-eutectic composition uniformly due to stronger convective mixing (Figure 5(c)). In water-24 wt% $NH_4Cl$, the liquid composition was much slower in reaching near-eutectic composition, and therefore the pure eutectic solid was considerably lesser. Moreover, the rate of growth of the uppermost location of the freezing interface was found to be considerably larger in the eutectic growth for the faceted case.

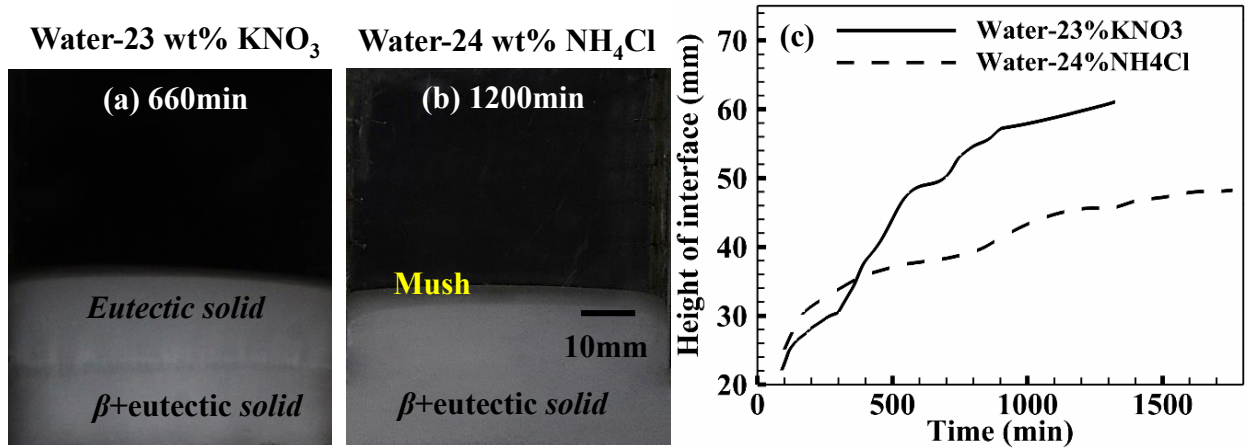

Figure 5: Evolution of solid (captured using Canon 1200D) during solidification of (a) water-23 wt% KNO$_3$ at 660min (b) water-24 wt% NH$_4$Cl at 1200min. (c) Comparison of the height of interface with the time of experiments between water-23 wt% KNO$_3$ and water-24 wt% NH$_4$Cl.

The existence of a fully eutectic phase and the mixture of primary and eutectic phases were ascertained by measuring the composition of the solid at the end of the experiments. The change in the mass of the solid sample was evaluated by evaporating the water from the solid mixture. In the faceted case, the average composition of β+eutectic and the eutectic was 55±3 wt% and 14±1 wt% KNO$_3$ respectively, whereas in the dendritic case, the average composition of the β+eutectic solid was 32±2 wt% NH$_4$Cl.

### 3.3 Velocity field in bulk fluid

In the faceted growth, the maximum velocity magnitude was 4.1 mm/sec at 100min (Figure 6(a)), and it reduced to 3.3 mm/sec (at 120min) with time (Figure 6(b)). As the solidification progressed, the convection strength decreased due to the shifting of composition towards the eutectic composition, which led to a decrease in velocity magnitude with time (Figure 6(a, b)).

After the initial solutal convection during dendritic growth, stabilized plumes were observed in the cell (Figure 6(c)). The approximate average velocity of plumes was 0.6 mm/s and that of the bulk fluid was much lower (~$10^{-2}$ mm/s). However, in the faceted case, the velocity field in the bulk fluid was uniform but larger in magnitude (2-4 mm/s).

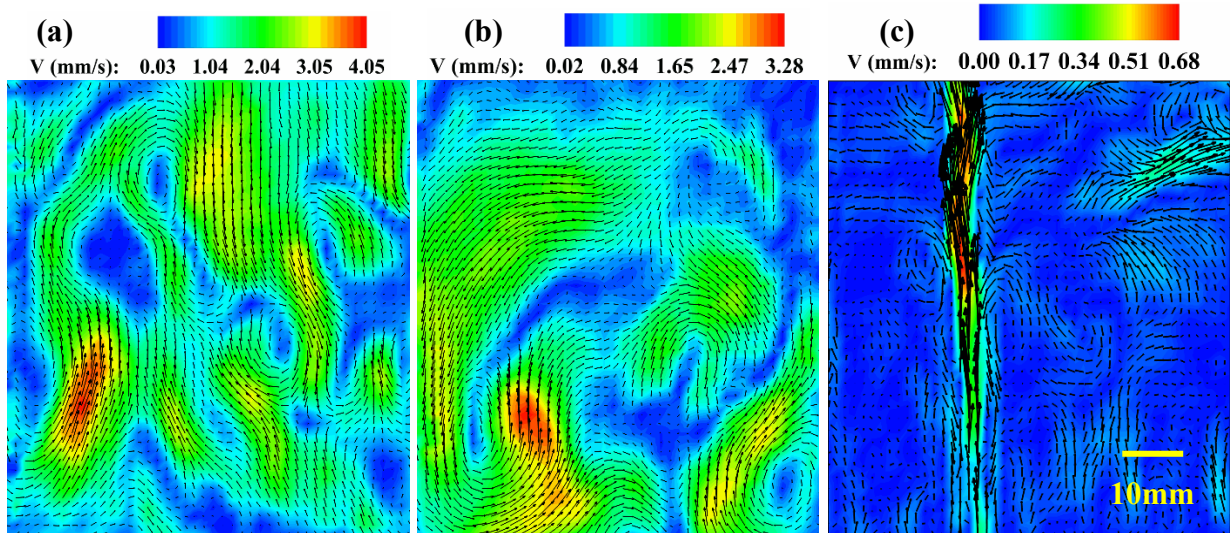

Figure 6: Velocity contours and vector in: faceted growth (water-23wt% $KNO_3$) at (a) 100min (b) 120min where random and higher velocity magnitude was observed compared to the stabilized plumes in dendritic growth (water-23wt% $NH_4Cl$) at 100min (c).

### 3.4 Uncharacteristic temperature rise in the faceted case

Figure 7 shows the temperature measurements at different locations along with the height of the cell for the faceted (water-23 wt% $KNO_3$, Figure 7(a)) and dendritic (water-24 wt% $NH_4Cl$, Figure 7(b)) growth cases. Significantly dissimilar temperature profiles were observed for the two cases, with the faceted solidification showing drastic initial cooling, followed by an unusual temperature rise. On the contrary, the temperature profiles of the dendritic case showed slow but gradual temperature decrease, on par with the nominal cooling rate. Moreover, both the cases demonstrated a slight temperature rise towards the end of the experiment (marked by the gray box in Figure 7(a-b)), which is very likely to have been associated with the heat gained from the surroundings. Based on the changes in the slope of temperature vs time measurements of the faceted case, the following three modes of heat transfer in the liquid have been identified: (a) convection-dominated regime, (b) transition regime, and (c) conduction-dominated regime. Note that similar regimes were identified in the dendritic case as well, but the changes in slopes were relatively less severe.

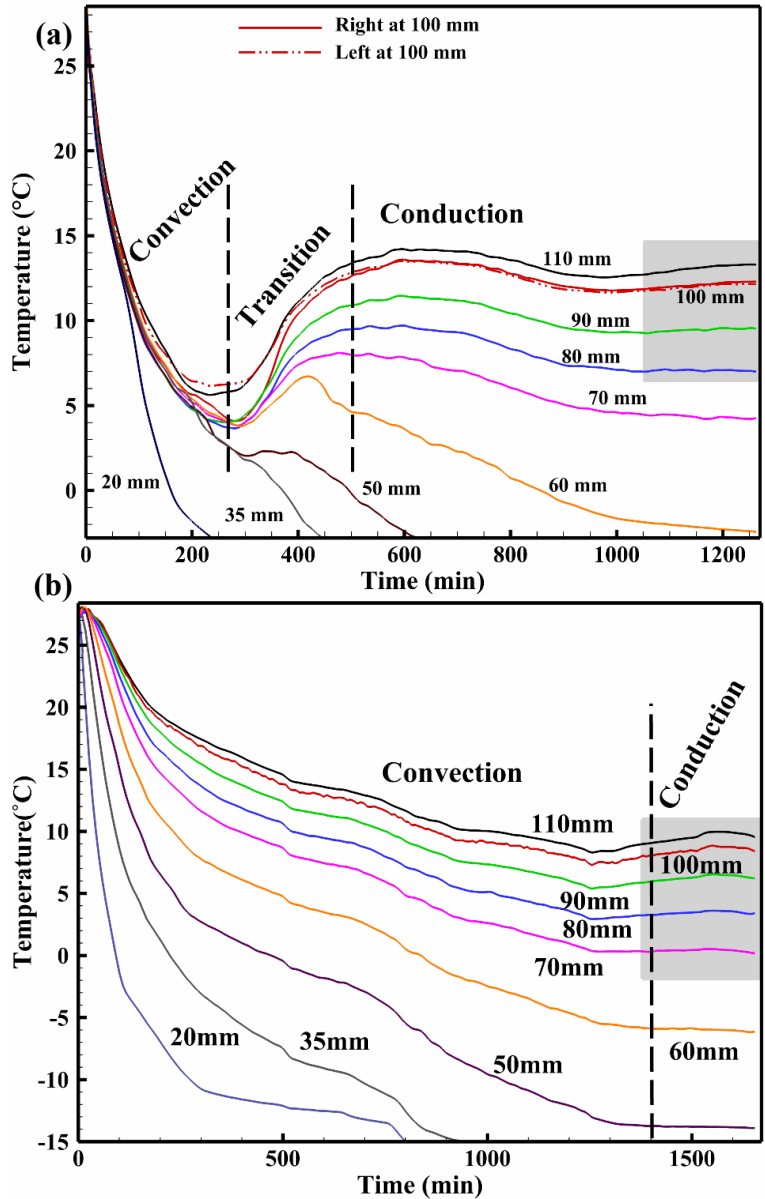

Figure 7: Temperature measurement at the different positions in the cell for (a) higher cooling rate, gain in temperature and conduction dominated cooling in the liquid was observed in solidification of water-23 wt% $KNO_3$ (b) slow cooling in liquid was observed in water-24 wt% $NH_4Cl$.

### 3.4.1 Convection-dominated regime

In the faceted case, vigorous and random convective flow decreased the temperature very rapidly and considerably weak vertical thermal gradients were noticed at all the measurement locations (Figure 7(a)). This phase is termed as the convection-dominated regime. For the dendritic case, the temperature of the fluid decreased with the nominal cooling rate (much lower than water-KNO$_3$ case) due to the presence of well-defined convective patterns (plumes), as shown in Figure 7(b).

### 3.4.2 Transition regime

At the end of the convection dominated regime (the magnitude of convective velocity can be seen from Figure 8(a), where the velocity is very low compared to Figure 6(a, b)), low thermal gradients were observed in the bulk fluid. The bulk fluid composition reached closer to the eutectic composition, where localized solutal convection (the local convective flow was in the order of 0.01 mm/s, as shown in Figure 8(b)), eutectic growth, and a gain in the temperature was observed. This region was termed as the transition regime (300min to 500min) (Figure 7(a)). Similar gain in the bulk liquid temperature was observed by Thompson et al. [23] during bottom cooled solidification of water-KNO$_3$-NaNO$_3$ where KNO$_3$ was the primary solidifying component. However, the increase was attributed to the heat gain from the surrounding environment.

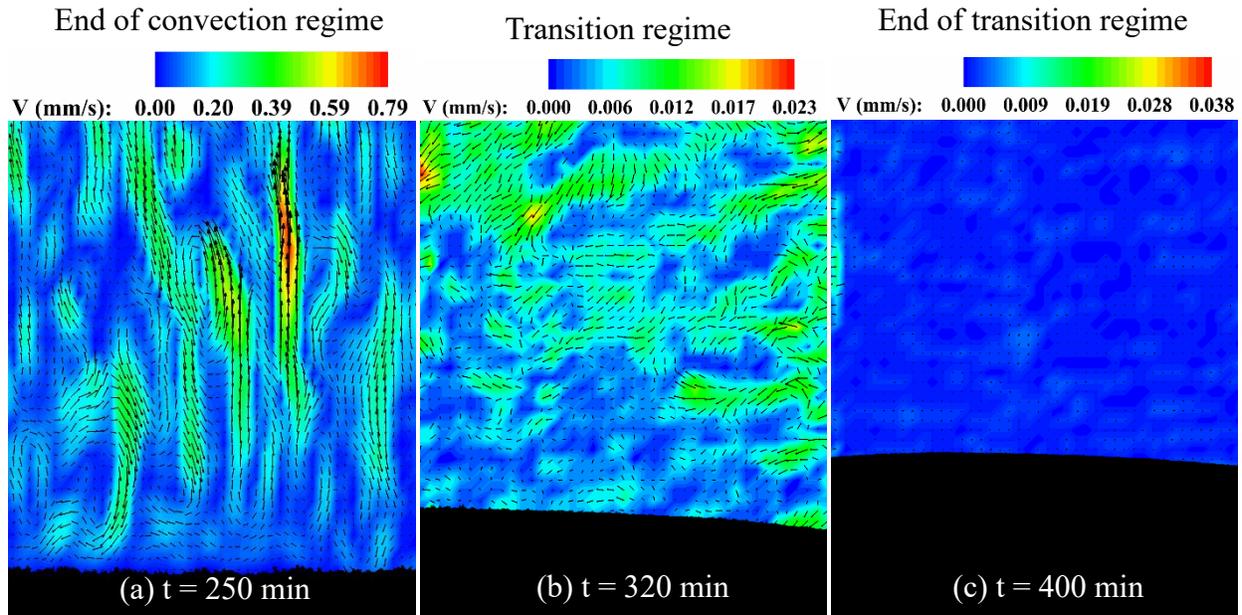

Figure 8: Velocity vectors and contours (a) at the end of the convection dominated regime, (b) at the start of the transition regime, (c) at the end of the transition regime.

In the present work, we further observed gain in the temperature with different initial compositions of water-KNO$_3$ system (experimental details of water-(15 and 20) wt% KNO$_3$ is presented in the appendix (figure A1 and A2)). In order to verify the rise in temperature in different faceted-forming mixtures, experiments with the hyper-eutectic mixtures of water-Na$_2$SO$_4$ were performed, where similar random convection patterns and gain in temperatures were observed during bottom-cooled solidification (data in appendix, figure (A3 and A4)). At the end of transition regime, the solid-liquid interface was flat, and beyond this regime, the phenomenon was considered to be conduction-dominated. In water-$NH_4Cl$ (dendritic growth) case, a positive thermal gradient in the liquid was available (Figure 7(b)). The gain in temperature, and hence the transition regime did not exist, unlike the cases of water-$KNO_3$ and water-$Na_2SO_4$ (faceted growth) systems.

### 3.4.3 Conduction-dominated regime

In the conduction-dominated regime, the temperature deceased gradually, with continuous growth of the eutectic solid (no convective flow was observed, as shown in Figure 8(c)). As a result, a larger thermal gradient in the liquid was set-up at the interface, which appeared to deflect the collimated light in the shadowgraphs by showing a high grayscale intensity. The intensity measurements at different locations near the solid-liquid interface (marked as rectangular box in Figures 3 and 4) are shown in Figure 9. A high intensity (grayscale intensity magnitude ~255) uniform band was observed in Figure 9(b) at the interface (740 min), unlike Figure 9(a) when the localized convection was present at 355min.

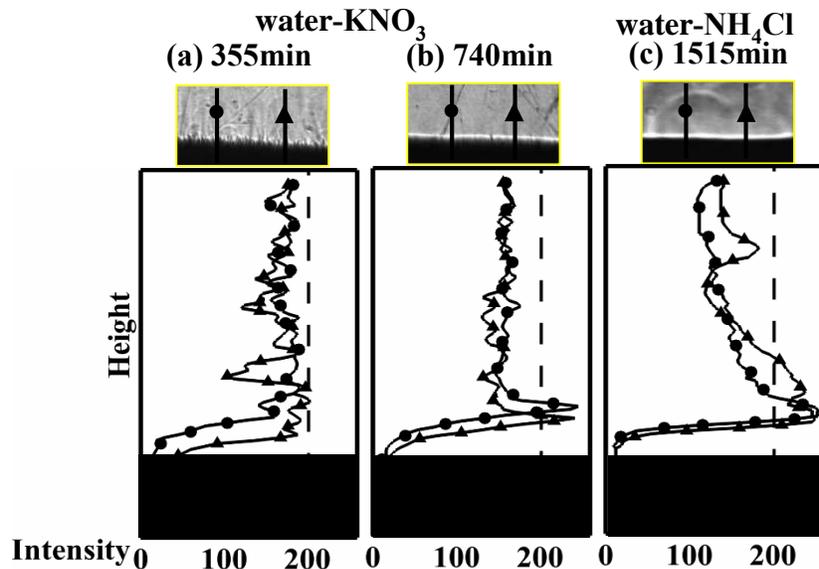

Figure 9: Correlation of the temperature gradients along with the height at the solid-liquid interface with the help of gray-scale intensity variation; (a) convection-dominated regime and (b) conduction-dominated regime in water-23 wt% $KNO_3$, (c) conduction-dominated regime in water-24 wt% $NH_4Cl$.

Similarly, in water-$NH_4Cl$ (Figure 4(d-f)) case, the eutectic growth led to a high gradient region at the interface after 1410 min. From the temperature and the intensity measurements, the gradient near the interface was observed to be higher in water-NH$_4$Cl (Figure 7(b), 9(c)) than the water-KNO$_3$ (Figure 7(a) and 9(b)) system.

## 4. Discussion

### 4.1 Analysis of flow behaviour using Rayleigh number

It is now clearly understood that the convective flow patterns in the bulk are strongly influenced by the mushy zone morphology. Based on two distinct volumetric zones of the flow, the following two cases were considered for the estimation of convective strength; (a) permeable or mushy zone fluid (b) bulk fluid.

#### 4.1.1 Natural convective flow through the permeable mushy zone

The strength of solutal convection in the mushy zone can be estimated using the permeability-based Rayleigh number ($Ra_{C\_P}$), which is formulated in Eq (1) [9,11].

$$Ra_{C\_P} = \frac{g\beta_C \Delta C_P Kl}{Dv} \qquad (1)$$

where $g$ is the gravitational acceleration, $\beta_C$ is the solutal expansion coefficient, $K$ is the average permeability, $l$ is the length scale, $D$ is the solutal diffusivity, $v$ is the kinematic viscosity, and $\Delta C_P$ is the composition difference between the initial and inter-granular regions. All the thermo-physical properties were taken as constant, and details are mentioned in (Table 1). $Kl$ is the volumetric length scale in the mushy zone. For unidirectional solidification, the growth rate is not constant; hence, the length scale ($l$) was chosen to be the height of the mushy zone instead of the ratio of thermal diffusivity and growth rate. The permeability based Rayleigh number ($Ra_{C\_P}$) was calculated to be 68 and 231 for the water-KNO$_3$ and water-NH$_4$Cl systems at 100 min, respectively. Literature suggested that below the critical permeability-based Rayleigh number ($Ra_{C\_P}$), the plumes were suppressed. In the present study, critical $Ra_{C\_P}$ was estimated to be 115. The critical $Ra_{C\_P}$ was evaluated with the minimum composition difference required at the onset of plumes.

The average permeability ($K = 6 \times 10^{-4} \lambda^2 \frac{(1-f_s)^3}{f_s^2}$, Blake-Kozeny equation [11]) is a function of the primary arm spacing ($\lambda$) and the mean solid fraction ($f_s$) in the mushy zone. These parameters were estimated from the images of the microstructures, shown in the insets of Figures 3(b) (for water-KNO₃) and 3(e) (for water-NH₄Cl). In the faceted case, The higher solid fraction and primary arm spacing typically result in a low value of the permeability (*K*) (Table 2), which, along with a lower mushy zone height (l) leads to a low solutal Rayleigh number. In the present study, it was observed that the permeability-based Rayleigh number was lesser for faceted cases than that of the critical $Ra_{C\_P}$ (Table 2), and therefore, the condition for the formation of the plumes was not met (Figure 3(a, b)).

Table 2: Experimental details during solidification of water-23wt% KNO₃ and water-24wt% NH₄Cl

| Parameters (Units) | Symbol | water-KNO₃ | water-NH₄Cl |
|---|---|---|---|
| Fraction of solid in mushy zone | $f_s$ | 0.82 | 0.45 |
| Primary arm spacing (mm) | $\lambda$ | 3 | 0.9 |
| Mushy zone height (mm) | $l$ | 11 | 19 |
| Average permeability at 100 min | $K$ | 4.7×10⁻¹¹ | 4.9×10⁻¹⁰ |
| Growth rate (μm/s) at 100 min | $V$ | 1.8 | 1.3 |
| Mushy zone length scale (m) | $\frac{\alpha}{V}$ | 0.08 | 0.1 |
| Growth rate (μm/s) | $V$ | 1.55 | 0.58 |
|  |  | (300-500 min) | (1000-1400 min) |

### 4.1.1 Natural convection flow in the bulk fluid

The convective velocity scale in the bulk fluid ($V_{scale} \simeq \sqrt{g\beta_C \Delta C h} \sim Ra_{C\_Bulk}^{0.5} h^{-1}$) is proportional to the solutal expansion coefficient ($\beta_C$), $\Delta C$ (composition difference between the initial and the instantaneous bulk fluid compositions in water-KNO₃ system, and composition difference between the initial and the plume compositions in water-NH₄Cl case) and *h* (height of liquid) or solutal *Ra* in bulk fluid ($Ra_{C-Bulk}$). For estimating the bulk fluid velocities, a differently scaled *Ra* is necessary.

The solutal Rayleigh number in bulk fluid ($Ra_{C\_Bulk}$) is defined with liquid height as length scale, and it is expressed as Eq. (2).

$$Ra_{C\_Bulk} = \frac{g\beta_C \Delta C h^3}{D\nu} \qquad (2)$$

where $h$ is the height of liquid and $h^3$ being the volumetric length scale for bulk fluid. Higher values of $\beta_C$, $\Delta C$ and $h$ (Figure 5(c)) led to higher solutal Rayleigh number in the faceted case (water-$KNO_3$). At 100 min, solutal Rayleigh number in the bulk fluid for water-23 wt % $KNO_3$ was $9.1\times10^9$ and $1.7\times10^9$ for water-24 wt % $NH_4Cl$. Similarly, water-$KNO_3$ has higher velocity scale (7.5 mm/s at 100 min) than water-$NH_4Cl$ (2 mm/s at 100 min), which agreed well with the experimental data obtained from PIV (Figure 6). Figure 10 shows the transient variation of non-dimensional solutal Rayleigh numbers in bulk fluid. Non-dimensional solutal Rayleigh number in bulk fluid is the ratio of instantaneous Rayleigh number to the maximum Rayleigh number. While the maximum solutal convection occurred at the same instant for both the cases, a lower mushy zone permeability of the faceted rapidly suppressed the convection.

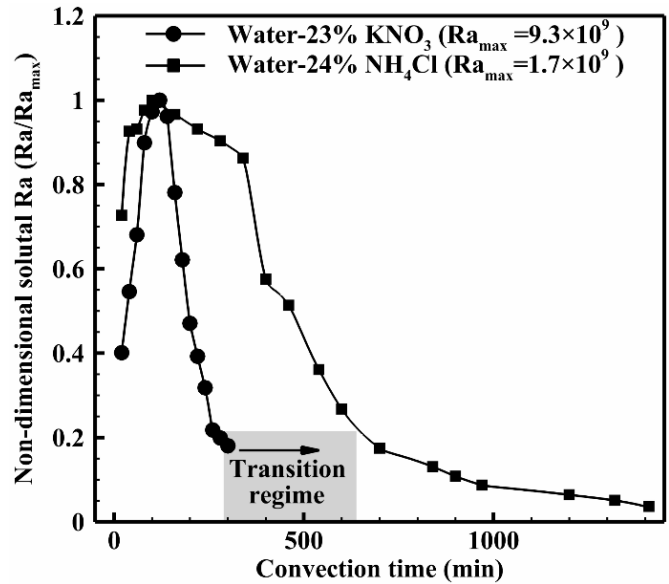

Figure 10: Non-dimensional solutal $Ra$ ($Ra/Ra_{max}$) in bulk fluid vs. time for the water-23wt% $KNO_3$ and water-23wt% $NH_4Cl$ experiments.

### 4.2 Energy balance in the transition regime

In order to correlate the role of natural convection patterns to the anomalous rise of liquid temperature in the faceted case, energy balance for a control volume was performed by considering the solidification of a differential volume element $dV$. Figure 11(a) shows an image of the experiment (water-KNO3, time = 400 min) in which a representative control volume at the interface is considered (Figure 11(b)). The energy balance at the interface for the growth of $dV$ amount of solid during the 300-500 min is expressed as Eq. (3)

$$k_s \frac{dT_s}{dx} - \rho_s HV = \rho_l C_{pl} \Delta T_{change} V + k_l \frac{dT_l}{dx} \qquad (3)$$

where $\frac{dT}{dx}$ is the thermal gradient, with subscripts $s$ and $l$ corresponding to the solid and liquid sides of the control volume, $k$ is the thermal conductivity, $\rho$ is the density, $H$ is the latent heat of fusion, $V$ is the average growth rate during 300-500min, $C_p$ is the specific heat capacity, $\Delta T_{change}$ is the change in temperature in liquid, $\Delta t$ is the duration of time (300-500 min). Thermo-physical properties and experimental details for the present study are shown in Table 1 and Table 2.

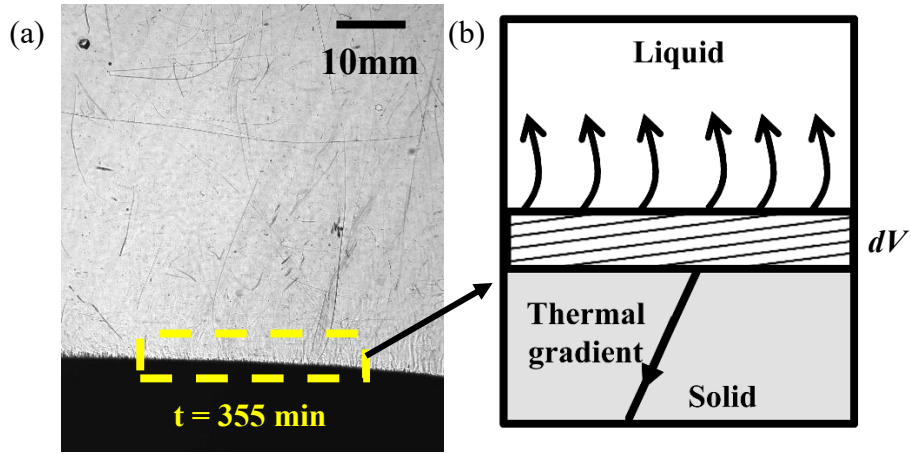

Figure 11: Energy balance at the liquid-solid interface in transition region (a) during faceted growth at 355min (b) schematic control volume.

It was assumed that the interface temperature of the growing solid was eutectic temperature, and the solid has uniform gradient. The gradient in the solid and the average solid-liquid interface position was taken at approximately 400 min (Figure 5(c)). In the presence of a negligible thermal gradient in the liquid due to random natural convection in the faceted case, a $\Delta T_{change}=13.1°C$ can be estimated from Eq. (3), which is approximately equal to the experimental gain of temperature in the liquid (in the transition regime, Figure 7(a)). However, for the dendritic case wherein the thermal gradient in the liquid was approximately 530°C/m, the estimated $\Delta T_{change}$ was approximately 1.2°C during a time interval of 1000-1400 min.

Based on these observations, a relation for estimation of an approximate order of the gain in temperature is given by the following relation:

$$\Delta T_{change} \approx \frac{\left(k_s \frac{dT_s}{dx} - \rho_s HV\right)}{\rho_l C_{pl} V} \quad (4)$$

After the reduction of local solutal convection, the composition ahead of the interface reached near eutectic. As the eutectic solid does not possess a mushy zone, a larger fraction of the solid was formed in the transition region compared to the previous regime. In the transition regime, a weak thermal gradient in the liquid was developed. The time scale for the development of the thermal gradient in eutectic growth was $\frac{\alpha}{V^2}$ (~$10^5$ sec), which is larger than the time duration associated with the transition regime (300-500min~ $10^4$ sec). This comparison validates the hypothesis that the gain of temperature in the liquid is primarily due to the reduced transport of heat in the solid.

The analysis finds that the uncharacteristic temperature rise during the solidification of the Water-$KNO_3$ mixture can be attributed to the localized solid fraction and morphology, controlled by the natural convection patterns ahead of the mushy zone. Though a length-scale such as the primary spacing was offered by the faceted morphology, the absence of secondary arms as that of the dendritic case meant that the liquid flow was completely restricted in the mushy zone. This was further aided by a higher solid fraction and early eutectic growth. Therefore, the fluid mixtures that freeze with low mushy zone permeability can exhibit a significant temperature increase during freezing. This study reveals the underlying physical mechanism behind the apparent heating of the liquid, which was previously thought to be associated with the laboratory surroundings.

## 5. Generalization of gain in temperature: A numerical study

A numerical model to simulate solidification was developed with an aim to generalize the plausible mechanism behind the gain in the liquid temperature. The aim of the simulation is to demonstrate the evolution of the thermal field during the onset of eutectic phase growth. A one-dimensional simulation of solidification was performed with the following assumptions, (a) the solid-liquid interface is flat during eutectic growth (this assumption is well justified by the experimental data presented earlier) (b) the system is in a bottom cooled configuration with a fixed temperature at the bottom and adiabatic top surface. The conditions of the simulation are based on the actual experimental measurements, to enable the generalization in both high and low Pr number fluids. The simulation methodology involves explicitly solving the heat equation with a Dirichlet type temperature boundary at the bottom, an adiabatic boundary at the top, and the Stefan condition at the solid-liquid interface. The heat equation for solid and liquid and the boundary conditions are given by Eq. (5)

$$\rho C_p \frac{\partial T}{\partial t} = \frac{\partial}{\partial x}\left(k \frac{\partial T}{\partial x}\right) \qquad 5(a)$$

$$T\big|_{x=0} = T_{Bottom} \qquad 5(b)$$

$$\frac{\partial T}{\partial x}\bigg|_{x=L} = 0 \qquad 5(c)$$

$$k_s \frac{dT_s}{dx}\bigg|_{x=x^*} - \rho HV = k_l \frac{dT_l}{dx}\bigg|_{x=x^*} \qquad 5(d)$$

where $\rho$ is the density, $t$ is the time, and $x^*$ is the position of the solid-liquid interface.

In systems that experience dendrite formation, convection in the liquid is literally non-existent at the onset of the formation of the eutectic phase. Thus, the system can be assumed to solidify under purely conductive heat transfer conditions, which is well described by the above model. However, systems that exhibit faceted growth experience almost negligible thermal gradients in liquid phase due to vigorous random convection in the bulk liquid prior to the onset of eutectic growth, which gradually reduces with time through the transition regime. The heat transfer in this regime is through conduction and localized random convection. The convective effects enhance the net heat flow into the liquid phase. Therefore, the net heat flow into the liquid is considered to be higher than that of the pure conduction alone, by approximately 2-4 times. This is justified by considering the orders of magnitudes of conductive (~20 W/m$^2$) and convective (~80 W/m$^2$) heat flux on the liquid side of the interface. Note that this exercise was performed with a primary objective of capturing an order of magnitude of the temperature change, with the help of a simplified one-dimensional model. For accurate predictions of temperature change, more exact, three-dimensional numerical simulations would be performed, which is not included to restrict the focus of the present investigation.

At the end of the transition regime, the system experiences a purely conductive heat transfer. Simulations were performed on NH$_4$Cl and KNO$_3$ materials (Figure 12). Further, simulations were also performed on low Prandtl number fluids such as alloys of aluminum (dendritic) and silicon (faceted), in order to observe the corresponding thermal fields (Figure 13).

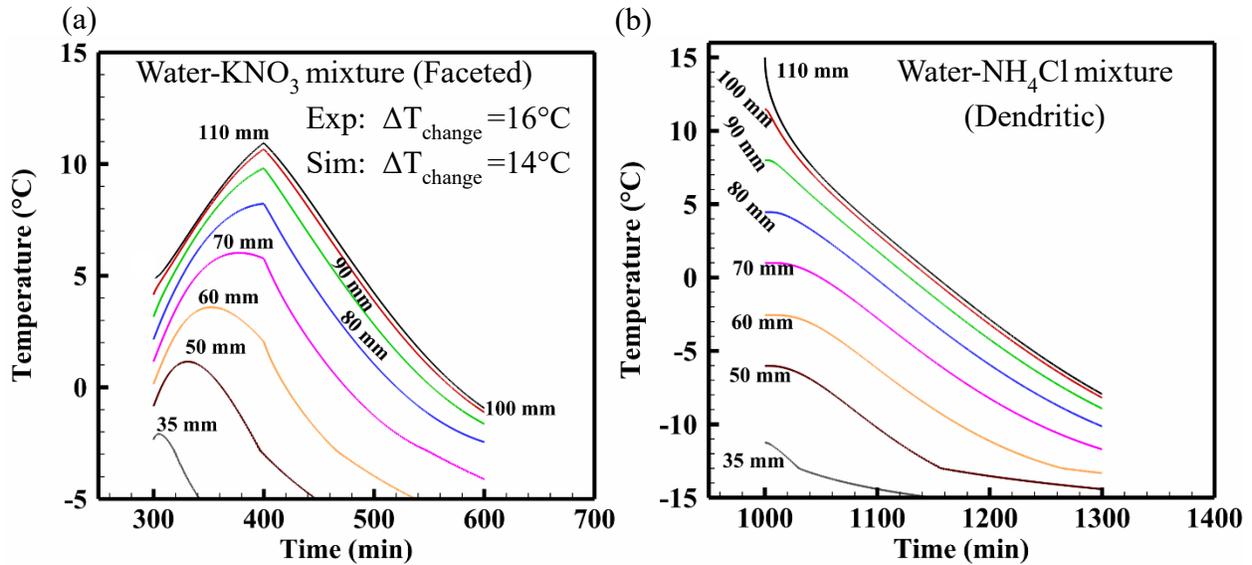

Figure 12: Temperature plot at different positions in the cell after the end of convection during (a) faceted growth of water-KNO$_3$ system, (b) dendritic growth in water-NH$_4$Cl system.

The temperature history at various locations along the height is evaluated, mimicking the experimental measurement locations. The comparison shows a clear agreement between the two sets of results (Figures 7 and 12). The NH$_4$Cl (dendritic) case predicts monotonically decreasing temperature with time, while the KNO$_3$ (faceted) case predicts a gain in the temperature profile in the transition regime. The extent of temperature rise is of similar magnitude as observed in the experimental results. Therefore, the mechanism of random convective flow causing an earlier eutectic transformation resulting in an increased liquid temperature is also ascertained with this numerical analysis. Similarly, numerical simulations on low Prandtl number fluids such as alloys of aluminium and silicon also predicted a similar temperature profile (Figure 13), thus indicating that the hypothesis could be extended to the realm of low Pr fluids.

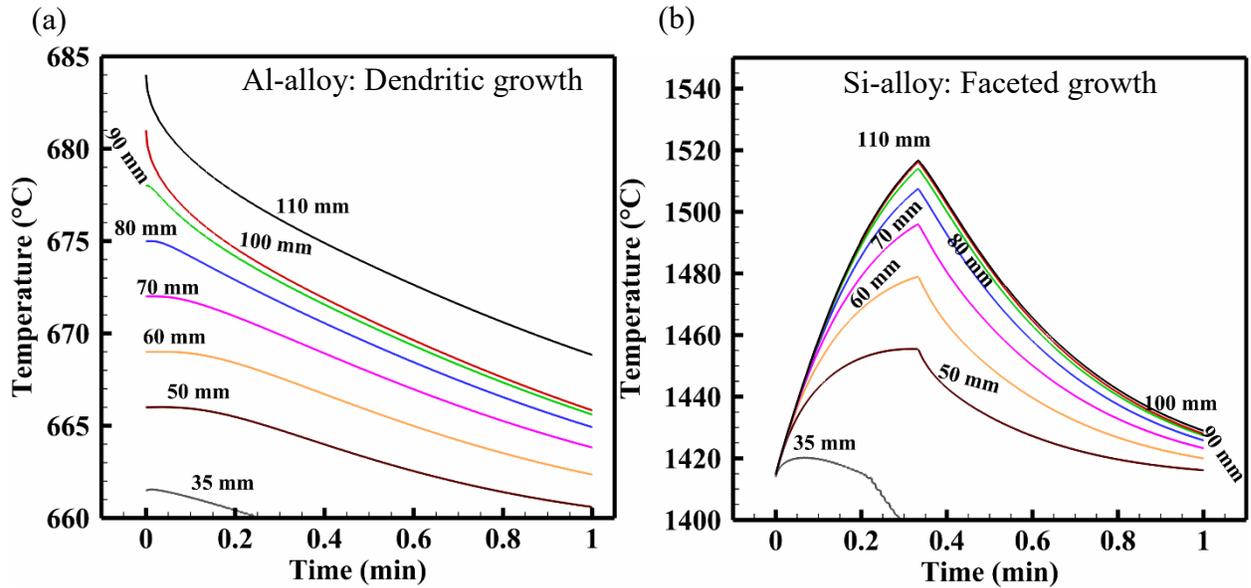
Figure 13: Temperature plot at different positions in the cell in the metallic system during solidification of (a) faceted growth in Si alloy system (b) dendritic growth in the Al alloy system.

**Conclusions**

The present experimental study investigated the role of microstructural morphology in governing the natural convection patterns in faceted and dendritic growth during bottom cooled solidification. Real-time visualization of the flow field was presented with the help of shadowgraph and PIV techniques. The investigation primarily sheds insights into the differences in mushy zone Rayleigh numbers between the dendritic and faceted cases, the nature of circulation in the bulk, and conclusively explains the reason behind an uncharacteristic gain in the bulk liquid temperature.

(i) The values of solutal Rayleigh number in the mushy zone ($Ra_{C\_P}$) were found to be 68 and 231, and whereas those in the bulk fluid Ra ($Ra_{C\_Bulk}$) were $9.1 \times 10^9$ and $1.7 \times 10^9$ for water-23 wt% $KNO_3$ (faceted) and water-24wt % $NH_4Cl$ (dendritic) respectively. The $Ra_{C\_P}$ for the faceted case was found to be considerably lower than the critical value that required for the plume formation (115).

(ii) As a result of the suppressed permeability, the convective flow patterns were predominantly in the form of plumes during dendritic cases, whereas completely random fluid motion in the bulk was observed in the faceted case. During dendritic solidification considerable thermal gradients continued to exist ahead of the interface throughout solidification resulting in a monotonic decrease of the bulk fluid temperature over time.

(iii) However, the faceted solidification was observed to develop an anomalous temperature rise in the transition region, at the onset of the eutectic solidification. This is attributed to the homogenization of the thermal fields in the liquid, resulting in a negligible thermal gradient ahead of the interface. This led to a much faster solidification and release of a larger amount of latent heat, which was further advected through the bulk liquid. The mechanism is further verified by changing the mixture composition and the mixture itself.

(iv) This mechanism is further generalized with the help of one-dimensional numerical model of solidification, in which the random convective mixing is simplified to be a homogenized temperature field. The analysis, performed on both high Pr (water-based) as well as low Pr (metallic) fluid systems exhibiting dendritic and faceted structures, ascertained the existence of an unconventional thermal behaviour during freezing.

The present work highlights the importance of mushy zone permeability, and has shed new insights into the role of natural convection in changing the bulk-fluid temperatures.

**Acknowledgment**

The authors gratefully acknowledge the financial support from the Department of Science and Technology (DST), India (Grant No. EMR/2015/001140).**Appendix**

Due to the high solid fraction in faceted growth (low permeability), the composition after the convection reaches the near eutectic composition in the early-stage of solidification (compared to the dendritic case) and experiences negligible thermal and compositional gradients in liquid. If the initial composition in the faceted cases is chosen to be near the eutectic, it may not be believable that due to random convection and the high solid fraction, the composition reaches eutectic. Hence the authors have chosen a water-23 wt% $KNO_3$, which is far from the eutectic. Additional observation of random convection, suppression of convection, and eutectic growth during solidification of water-15 wt% $KNO_3$ (Figure A1(a-c)) and water-18 wt% $KNO_3$ (Figure A1(d-f)) is reported in current section. Gain in temperature during transition regime during water-15 wt% KNO$_3$ (Figure A2(a)) and water-18 wt% KNO$_3$ (Figure A2(b)), which is similar to water-23 wt% KNO$_3$ (Figure 7).

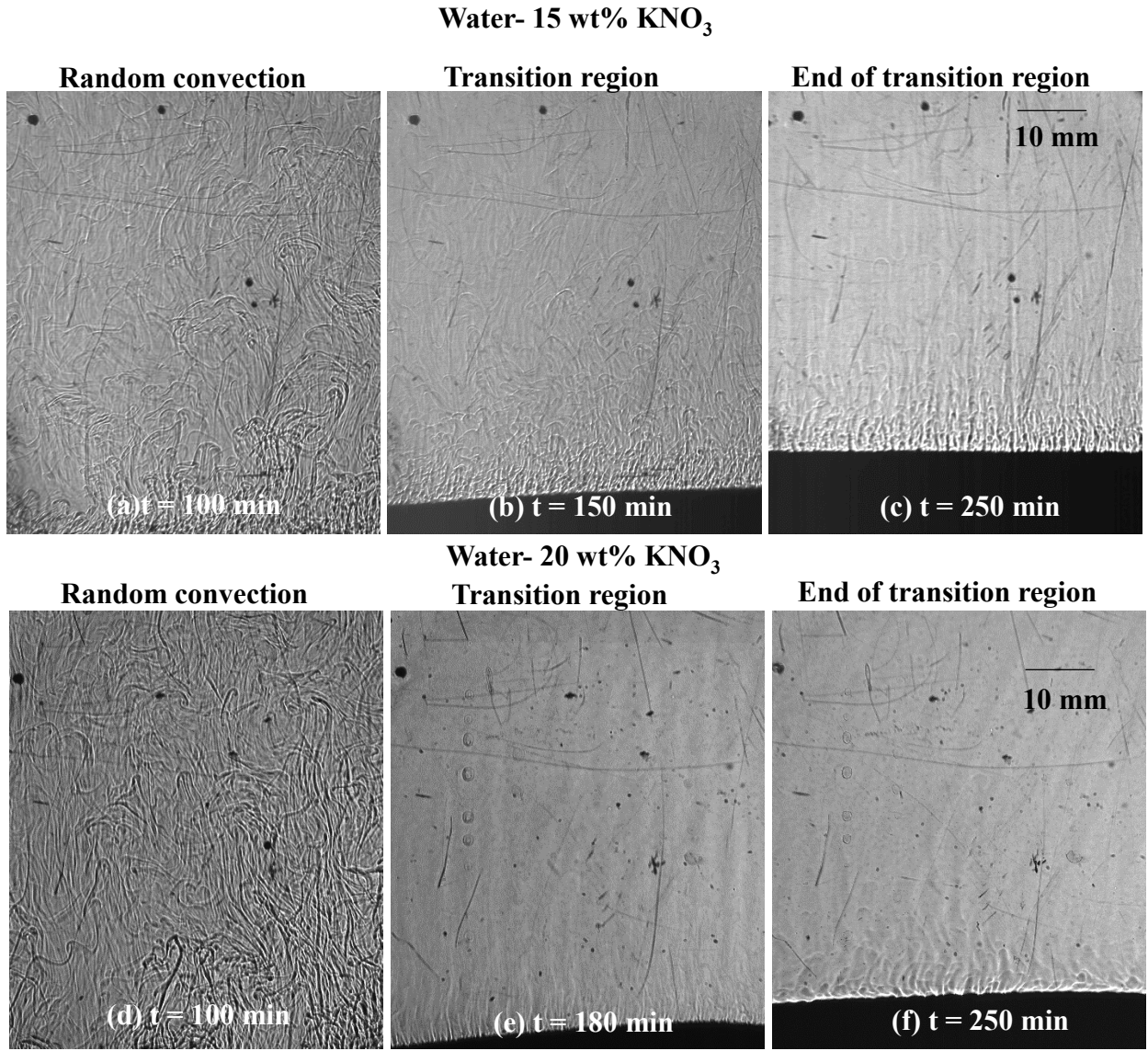

Figure A1: Convective flow observation using shadowgraph (a-c) for water-15 wt% KNO$_3$ (d-f) for water-20 wt% KNO$_3$: (a, d) random solutal convection at 100 min (b, e) reduction in convective strength and localized solutal convection (c, f) bulk composition near to eutectic and eutectic growth at 250 min.

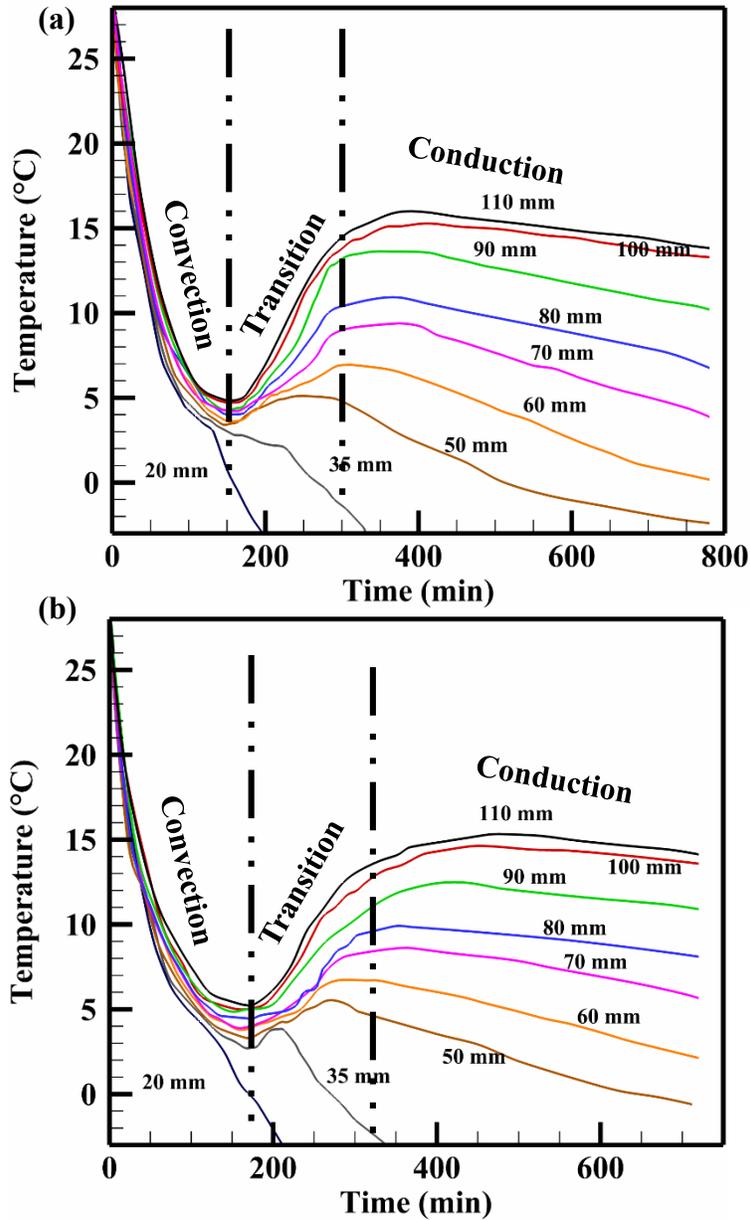

Figure A2: Temperature measurement at the different positions in the cell during solidification of (a) water-15 wt% $KNO_3$ (b) water-20 wt% $KNO_3$.

The observation of random convection patterns was further corroborated by performing experiments on a similar binary mixture that exhibits faceted morphology upon solidification. The faceted growth and random convection were observed during solidification of water-18 wt% $Na_2SO_4$ as shown in Figure A3, which were similar to water-23 wt% $KNO_3$. The temperature measurements show the three regimes are similar to the water-$KNO_3$ case, as shown in Figure A4. Note: Eutectic temperatures of binary eutectic of water-4 wt% $Na_2SO_4$ is -1.5 °C [32].

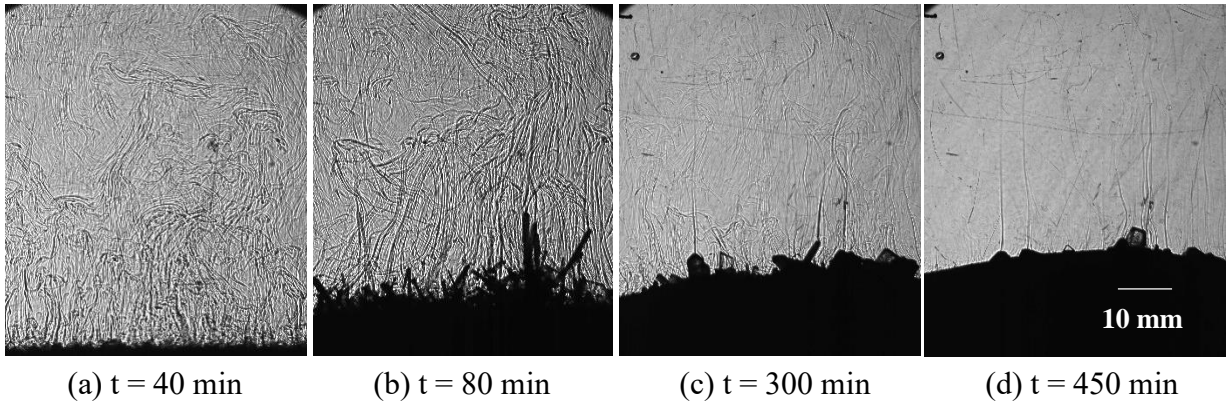

| (a) t = 40 min | (b) t = 80 min | (c) t = 300 min | (d) t = 450 min |

Figure A3: Convective flow observation using shadowgraph (a-d) for water-18 wt% $Na_2SO_4$: (a) random solutal convection at 60 min (b) similar flow at 260 min, and faceted microstructure was observed (c) reduction in convective strength (d) localized solutal convection and composition reached near eutectic composition and the sloped interface was observed at 450 min.

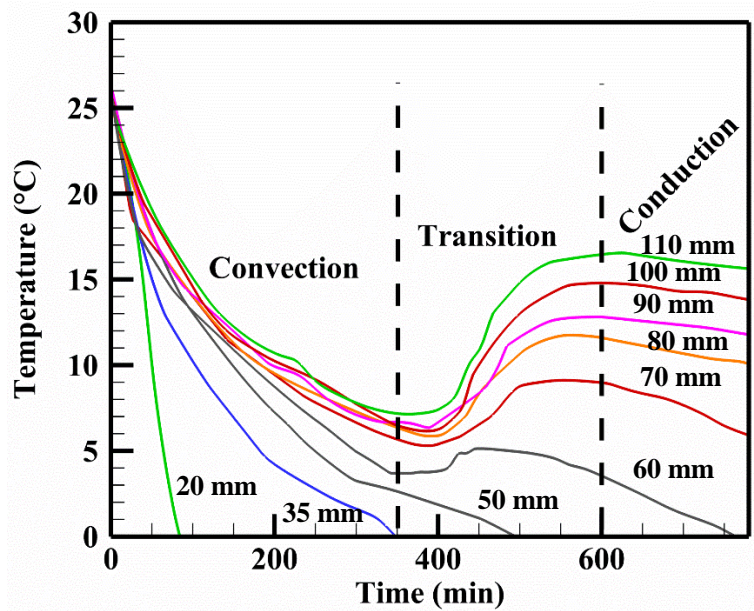

Figure A4: Temperature measurement at the different positions in the cell during solidification of water-18 wt% $Na_2SO_4$.

Observation of plume formation, suppression of plume and localized convection during solidification of water-27 wt% $NH_4Cl$ (Figure A5(a-c)). Figure A6 show the decreases in temperature at all the position in the cell during dendritic growth.

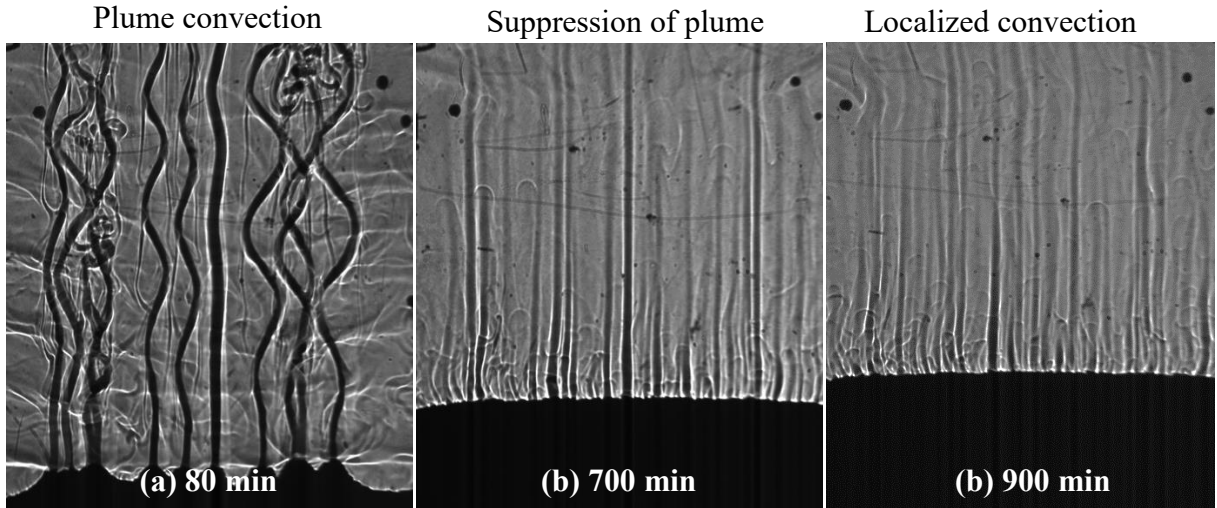

Figure A5: Convective flow observation using shadowgraph during solidification of water-27 wt% NH$_4$Cl.

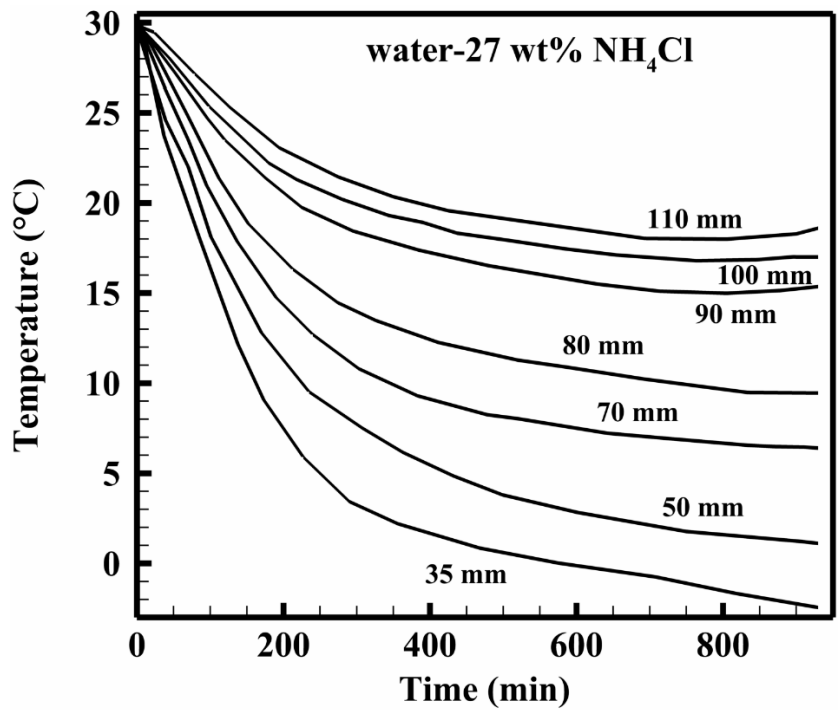

Figure A6: Temperature measurement at the different positions in the cell during solidification of water-27 wt% NH$_4$Cl. all the thermocouple temperatures decrease with nominal cooling rate in presence of dendrites.

# References

[1]　C. Beckermann, J.P. Gu, W.J. Boettinger, Development of a freckle predictor via rayleigh number method for single-crystal nickel-base superalloy castings, Metall. Mater. Trans. A. 31 (2000) 2545–2557. doi:10.1007/s11661-000-0199-7.

[2]　N. Shevchenko, S. Boden, G. Gerbeth, S. Eckert, Chimney formation in solidifying Ga-25wt pct In alloys under the influence of thermosolutal melt convection, Metall. Mater. Trans. a-Physical Metall. Mater. Sci. 44A (2013) 3797–3808. doi:DOI 10.1007/s11661-013-1711-1.

[3]　S.M. Copley, A.F. Giamei, S.M. Johnson, M.F. Hornbecker, The origin of freckles in binary alloys, Metall. Trans. 1 (1970) 2193–2204. papers2://publication/uuid/C23132B5-AA63-4868-B2A9-2AC831AB15F5.

[4]　L. Yuan, P.D. Lee, A new mechanism for freckle initiation based on microstructural level simulation, Acta Mater. 60 (2012) 4917–4926. doi:10.1016/j.actamat.2012.04.043.

[5]　S. Karagadde, L. Yuan, N. Shevchenko, S. Eckert, P.D. Lee, 3-D microstructural model of freckle formation validated using in situ experiments, Acta Mater. 79 (2014) 168–180. doi:10.1016/j.actamat.2014.07.002.

[6]　D.R. Poirier, Permeability for flow of interdendritic liquid in columnar-dendritic alloys, Metall. Trans. B. 18 (1987) 245–255. doi:10.1007/BF02658450.

[7]　M.C. Schneider, J.P. Gu, C. Beckermann, W.J. Boettinger, U.R. Kattner, Modeling of micro- and macrosegregation and freckle formation in single-crystal nickel-base superalloy directional solidification, Metall. Mater. Trans. A. 28 (1997) 1517–1531. doi:10.1007/s11661-997-0214-3.

[8]　S.D. Felicelli, D.R. Poirier, J.C. Heinrich, Modeling freckle formation in three dimensions during solidification of multicomponent alloys, Metall. Mater. Trans. B. 29 (1998) 847–855. doi:10.1007/s11663-998-0144-5.

[9]　M.G. Worster, Convection in mushy layers, Annu. Rev. Fluid Mech. 29 (1997) 91–122. doi:10.1146/annurev.fluid.29.1.91.

[10]　M.G. Worster, Natural convection in a mushy layer, J. Fluid Mech. 224 (1991) 335–359. doi:10.1017/S0022112091001787.

[11]　J.C. Ramirez, C. Beckermann, Evaluation of a rayleigh-number-based freckle criterion for Pb-Sn alloys and Ni-base superalloys, Metall. Mater. Trans. A. 34 (2003) 1525–1536. doi:10.1007/s11661-003-0264-0.

[12] D.A. Porter, K.E. Easterling, Phase transformations in metals and alloys, second edi, Springer-science Business Media,B.V., 1992.

[13] H.E. Huppert, The fluid mechanics of solidification, J. Fluid Mech. 212 (1990) 209–240. doi:10.1017/S0022112090001938.

[14] R.C. Kerr, A.W. Woods, H.E. Huppert, M.G. Worster, Solidification of an alloy cooled from above part 1. Equilibrium growth, J. Fluid Mech. 216 (1990) 323–342. doi:10.1017/S002211209000074X.

[15] I. Karaca, E. Çadirli, H. Kaya, N. Maraşli, Directional solidification of pure succinonitrile and succinonitrile- salol alloys, Turkish J. Phys. 25 (2001) 563–574.

[16] C.F. Chen, J.S. Turner, Crystallization in a double-diffusive system, J. Geophys. Res. 85 (1980) 2573–2593.

[17] R.C. Kerr, A.W. Woods, H.E. Huppert, M.G. Worster, Solidification of an alloy cooled from above part 2. Non-equilibrium interfacial kinetics, J. Fluid Mech. 217 (1990) 331–348. doi:10.1017/S002211209000074X.

[18] R.C. Kerr, A.W. Woods, H.E. Huppert, M.G. Worster, Solidification of an alloy cooled from above. Part 3. Compositional stratification within the solid, J. Fluid Mech. 218 (1990) 337–354. doi:10.1017/S0022112090001021.

[19] J.A. Dantzig, M. Rappaz, Analytical solutions for solidification, in: Solidifcation, EPFL Press, 1994: pp. 165–170.

[20] L.J. Bloomfield, H.E. Huppert, Solidification and convection of a ternary solution cooled from the side, J. Fluid Mech. 489 (2003) 269–299. doi:10.1017/S0022112003005172.

[21] F. Chen, Formation of double-diffusive layers in the directional solidification of binary solution, J. Cryst. Growth. 179 (1997) 277–286. doi:10.1016/S0022-0248(97)00098-5.

[22] V. Kumar, A. Srivastava, S. Karagadde, Compositional dependency of double-diffusive layers during binary alloy solidification: Full-field measurements and quantification, Phys. Fluids. 30 (2018) 113603. doi:10.1063/1.5049135.

[23] A.F. Thompson, H.E. Huppert, M.G. Worster, A. Aitta, Solidification and compositional convection of a ternary alloy, J. Fluid Mech. 497 (2003) 167–199. doi:10.1017/S002211200300661X.

[24] C. Beckermann, C.Y. Wang, Equiaxed dendritic solidification with convection: Part III. Comparisons with NH4Cl-H2O experiments, Metall. Mater. Trans. A. 27 (1996) 2784–

2795. doi:10.1007/BF02652371.

[25] A. Aitta, H.E. Huppert, M.G. Worster, Diffusion-controlled solidification of a ternary melt from a cooled boundary, J. Fluid Mech. 432 (2000) 201–217.

[26] D.G. Cherkasov, M.P. Smotrov, K.K. Il'in, Topological transformation of the phase diagram of the potassium nitrate-water-n-butoxyethanol ternary system, Russ. J. Phys. Chem. A. 84 (2010) 922–927. doi:10.1134/S0036024410060063.

[27] K. Tanaka, Measurements of self-diffusion coefficients of water in pure water and in aqueous electrolyte solutions, J. Chem. Soc., Faraday Trans. 1. 71 (1975) 1127–1131. doi:10.1039/F19757101127.

[28] P.R. Chakraborty, P. Dutta, Study of Freckles Formation During Directional Solidification Under the Influence of Single-Phase and Multiphase Convection, J. Therm. Sci. Eng. Appl. 5 (2013) 021004 (1–11). doi:10.1115/1.4023601.

[29] V. Kumar, A. Srivastava, S. Karagadde, Real-time observations of density anomaly-driven convection and front instability during solidification of water, J. Heat Transfer. 140 (2018) 042503. doi:10.1115/1.4038420.

[30] V. Kumar, M. Kumawat, A. Srivastava, S. Karagadde, Mechanism of flow reversal during solidification of an anomalous liquid, Phys. Fluids. 29 (2017) 123603. doi:10.1063/1.5005139.

[31] V. Kumar, A. Srivastava, S. Karagadde, Do the intrusive probes alter the characteristic length-scales of natural convection?, J. Flow Vis. Image Process. 25 (2018) 207–228.

[32] C. McCarthy, R.F. Cooper, S.H. Kirby, K.D. Rieck, L.A. Stern, Solidification and microstructures of binary ice-I/hydrate eutectic aggregates, Am. Mineral. 92 (2007) 1550–1560. doi:10.2138/am.2007.2435.